# Momentum selectivity and anisotropy effects in the nitrogen K-edge resonant inelastic X-ray scattering from GaN

V.N. Strocov[1,*], T. Schmitt[2,*,#], J.-E. Rubensson[2], P. Blaha[3],
T. Paskova[4], and P.O. Nilsson[5]

[1]Paul Scherrer Institute, CH-5232 Villigen PSI, Switzerland
[2]Department of Physics, Uppsala University, Box 530, S-75121 Uppsala, Sweden
[3]Institut für Materialchemie, Technische Universität Wien, A-1060 Wien, Austria
[4]Linköping University, S-581 83 Linköping, Sweden
[5]Chalmers University of Technology, S-412 96 Göteborg, Sweden

High-resolution soft X-ray emission and absorption spectra near the N *K*-edge of wurtzite GaN are presented. The experimental data are interpreted in terms of band structure based full-potential electronic structure calculations. The absorption spectra, compared with calculations including core hole screening, indicate partial core hole screening in the absorption process. The resonant emission spectra demonstrate pronounced dispersions of the spectral structures, identifying effects of momentum conservation due to resonant inelastic X-ray scattering (RIXS) with anisotropic electronic structure of GaN. In view of a wide range of optoelectronic applications of GaN, our findings on the momentum selectivity in RIXS can be utilized in development of GaN based nanoelectronics devices by controlling direct *vs* indirect band gap character of GaN nanostructures.

## I. INTRODUCTION

Soft-X-ray emission (SXE) and absorption (SXA) spectroscopies allow investigation of the electronic structure of the valence band (VB) and conduction band (CB), respectively, with elemental specificity and large probing depth up to 3000 Å. In particular, the electronic structure of buried quantum dots, interfaces[1] and even isolated impurities[2] (see further references in a review[3] by Kotani and Shin) can be accessed by these techniques.

In general, the SXE/SXA spectroscopies characterize the electron partial density of states (PDOS) resolved in their elemental and orbital character, but averaged over the wave vectors $\mathbf{k}$. For weakly correlated semiconductor systems, however, certain $\mathbf{k}$-selectivity appears in resonant inelastic X-ray scattering (RIXS) where the core electron is excited within a few eV above the absorption threshold.[3] In this case the absorption and emission events are coupled in a fast coherent scattering process, in which the full momentum is conserved as

$$\mathbf{q}_{in} - \mathbf{q}_{out} = \mathbf{k}_e + \mathbf{k}_h$$

where $\mathbf{q}_{in}$ and $\mathbf{q}_{out}$ are wave vectors of the absorbed and emitted photons, and $\mathbf{k}_e$ and $\mathbf{k}_h$ are those of the conduction electron and valence hole, respectively. Dependence of the resonant SXE spectra on the excitation energy $h\nu_{ex}$ reflects then the VB and CB dispersions $E(\mathbf{k})$. The $\mathbf{k}$-selectivity of RIXS has been demonstrated for a number of semiconductors such as Si and SiC, and even for the graphite semimetal.[3,4,5]

GaN is a prototype nitride compound representative of a large family of binary and ternary group III nitrides, having a wide range of optoelectronic device applications covering the entire region from 200 to 1200 nm. Presently commercial blue laser diodes are available, and lasers working in the blue-violet spectral region are realized for both pulsed and continuous wave operations. However, these significant achievements do not fully exploit the potential applicability of nitride compounds. Further progress can be expected due to low dimensional nitride devices like quantum dot lasers or single photon emitters utilising, for example, self-assembled GaN quantum dots in AlN or AlGaN matrix. The current strong interest in such systems is motivated by their potential to achieve lasers with lower threshold currents and improved temperature characteristics. Therefore, a better control of the properties of such nitride based nanostructures is of significant practical interest.

Spectroscopic investigations of the GaN electronic structure are far from being exhaustive. In particular, its $\mathbf{k}$-resolved studies with conventional angle-resolved photoelectron spectroscopy (ARPES) are burdened by surface effects sensitive to the sample growth and surface preparation procedures (see, e.g., Refs. 6 and 7). Moreover, small photoelectron escape depth principally limits applications of this technique to nanostructure systems such as GaN quantum dots buried in a matrix of other material. The soft-X-ray spectroscopies with their large probing depth and elemental specificity are ideally suited for this purpose. Stagarescu *et al.*[8] and Lawniczak-Jablonska *et al.*[9] used the SXE and SXA spectroscopies to study, respectively, the occupied and unoccupied PDOS of GaN,. Eisebitt *et al.*[10] found pronounced changes in the N *K*-edge SXE spectral shape occuring near the absorption threshold, which they interpreted in terms of momentum conservation. However, a systematic study of RIXS in GaN and its connection to the band structure is presently missing.

Here, we present a high-resolution SXE/SXA study on wurtzite GaN performed near the N *K*-edge. Compared to our pilot study in Ref. 11, the present results are based on experimental data obtained for a larger set of excitation energies at higher resolution and varied scattering geometries, and receive extended theoretical analysis. We unambiguously identify effects of momentum conservation in the RIXS process. The experimental results are supported by our state-of-the-art band structure calculations.

## II. EXPERIMENT

We have studied wurtzite GaN with the (0001) surface orientation. The sample was a 30 μm thick GaN film grown on a sapphire substrate using hydride vapour phase epitaxy in the optimum growth window.[12] High crystalline quality of the sample was confirmed by high resolution x-ray diffraction rocking curve and low temperature photoluminescence (LT-PL) measurements. The FWHM values of the ω- and 2θ-ω radial scans for the symmetric (002) and asymmetric (102), (104), and (114) reflections approached the best values reported for thick nitride films grown on foreign substrate, and indicate relatively low mosaicity without domain formation achieved at this thickness. The LT-PL spectrum of the sample shows well resolved exciton-related peaks in the near band edge region with FWHM less than 2 meV, confirming the high crystalline and optical quality of the film.

The SXE/SXA experimental data were taken near the N *K*-edge (1*s* core level) at ~400 eV. The experiments were performed at MAX-lab, Sweden, at the undulator beamline I511-3 equipped with an SX-700 type plane grating monochromator,[13] and a high-resolution Rowland-mount grazing incidence SXE spectrometer.[14] The incident radiation had linear *p*-polarization (**E** vector in the incidence plane). The sample was positioned at a grazing angle of ~20° relative to the incident beam.

The SXA data were measured with the monochromator resolution set to 0.2 eV, which is comparable to the N 1*s* core level intrinsic width of about 0.1eV.[9] The absorption signal was recorded in total electron yield mode and normalized to the photocurrent from a gold mesh introduced into the synchrotron radiation beam upstream the experimental chamber.

The experimental SXA spectrum is shown in Fig.1 (*right spectrum on top*). With the N 1*s* core level, it reflects the CB composed predominantly of the antibonding N 2*p* states. Due to high anisotropy of GaN the SXA spectra show significant angle dependence (see below). Our spectrum is in good agreement with previous SXA data[9,15] measured at similar grazing geometry, although with slight differences in the peak amplitudes due to a somewhat different incidence angle.

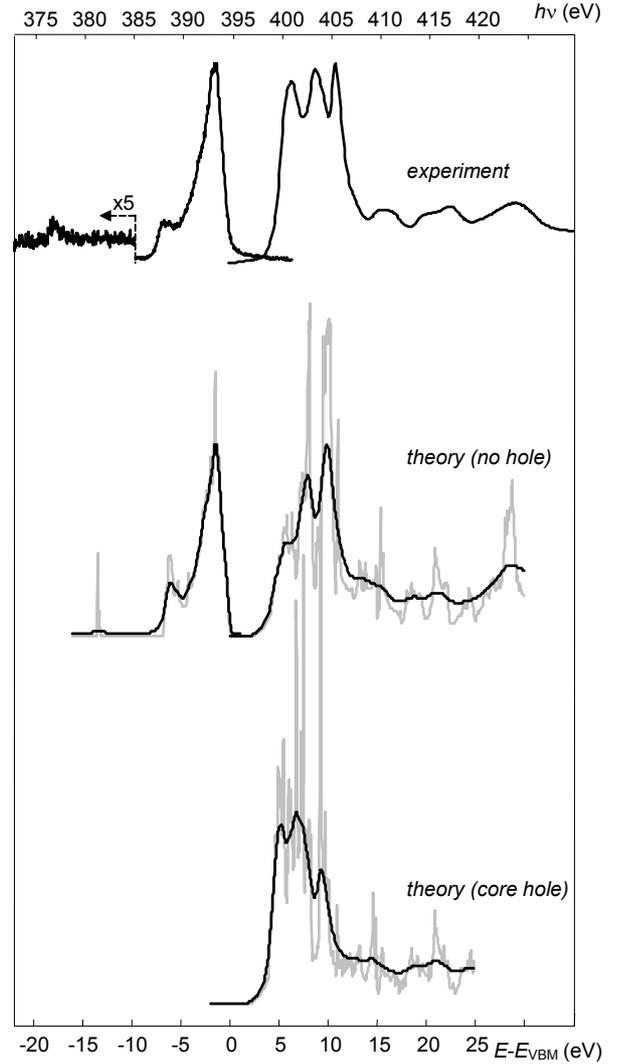

FIG. 1. (*top*) Experimental N 1*s* off-resonant SXE spectrum ($h\nu_{ex}$~420eV) and SXA spectrum, in comparison with (*gray curves below*) theoretical results excluding and including the core hole. The empirical simulation of the lifetime and instrumental broadening effects (see the text) is shown on top of the theoretical curves. With our grazing incidence experimental geometry, the spectra reflect mainly the N $p_{xy}$ PDOS in the VB and N $p_z$ PDOS in the CB.

The SXE measurements were performed with the spectrometer installed in the incidence plane at a scattering angle of 90° relative to the incident beam. SXE spectra were taken in 1st diffraction order with a spherical grating of 5 m radius and 1200 lines/mm groove density, providing a resolution around 0.2 eV. The energy scale of the spectrometer was set using the tabulated characteristic X-ray line energies of the Co $L_\alpha$ and $L_\beta$ lines from Ref. 16 recorded from a pure Co-foil in 2nd diffraction order. Based on the elastic peaks in the SXE spectra, this energy scale was then used to calibrate the monochromator relative to the spectrometer energy scale. For further experimental details see Ref. 2.

The experimental off-resonant SXE spectrum shown in Fig.1 (*top left spectrum*) is in good agreement with previous off-resonant data.[8] This spectrum, taken at an excitation energy of 428.7 eV, well above the N K-edge, was further used as a reference for the resonant series.

In the energy region from 387 to 395 eV, the SXE spectrum shows the main N $2p$ line, reflecting the VB composed predominantly of the bonding N $2p$ states, with two peaks. In addition, it shows a sizeable structure near 377.5eV, which first was noticed in Ref. 8 and interpreted as due to hybridization of the Ga $3d$ semicore states with N $2p$ states. However, we find this structure about 2 eV deeper in energy compared to Ref. 8. Due to better statistics and resolution we can also identify a high-energy shoulder in this structure. Dhesi et al.[6] in their ARPES study of GaN ascribed a similar spectral feature to N $2s$ states.[17]

The experimental resonant SXE spectra in comparison with the reference off-resonant one (at the bottom) are shown in Fig.2. The corresponding excitation energies $h\nu_{ex}$ in relation to the SXA structures (see inset of Fig. 2) are indicated on the right. The resonant spectra were measured with a monochromator resolution of better than 0.2 eV, matching that of the spectrometer. The resonant SXE series shows clear dispersions and lineshape changes of both spectral peaks in the VB, with their dispersion ranges being about 0.3eV for the lower-energy peak, and 1.0eV for the higher-energy one. This identifies pronounced effects of momentum conservation in the RIXS spectra of GaN. Interestingly, the onset of the Ga $3d$ derived structure with $h\nu_{ex}$ seems to be slightly delayed compared to the main N 2p line.

## III. COMPUTATIONS

Our computations of the band structure $E(\mathbf{k})$ and PDOS were performed within the standard DFT formalism. The electron exchange-correlation was described within the generalized gradient approximation (GGA).[18] The calculations employed a full-potential LAPW+local orbitals method[19,20] implemented in the WIEN2k package (Ref. 21).

The SXA and off-resonant SXE spectra were calculated within the same computational framework. The SXA calculations were performed, first, with complete neglect of the core hole within the 'initial-state approximation'[22] (for calculations including the core hole see below). The SXE calculations employed 'the final-state rule', where the core hole is filled.[23] Within these approaches the SXA/SXE spectra appear as the element and orbital projected PDOS, multiplied with the energy dependent dipole matrix elements between the core and the corresponding CB/VB states.[24] Since the wave function of the core-states are well localized inside the atomic spheres (defined in the APW+lo method), the integrals can be restricted to that region. These dipole matrix elements are usually quite smooth functions with energy and do not alter the corresponding PDOS too much. (This is different when one has a $p$-type core hole, e.g. an $L_{2,3}$ spectrum, where the contributions of the $p\rightarrow s$ and $p\rightarrow d$

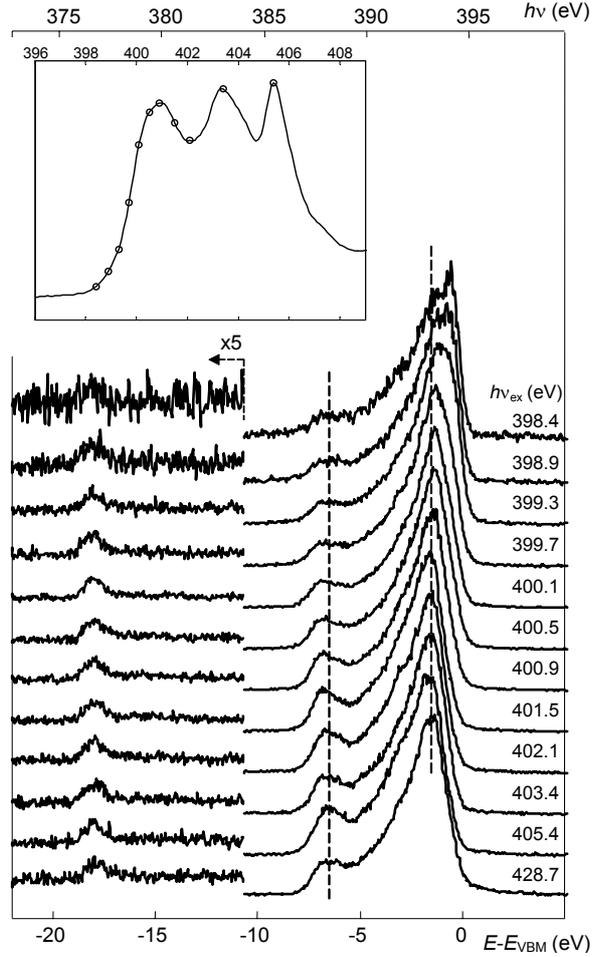

FIG. 2. Experimental N $1s$ resonant SXE series in comparison with the off-resonant spectrum (*bottom curve*), all normalized to the same peak value (relative to the foot background). The indicated excitation energies $h\nu_{ex}$ are also marked at the SXA spectrum (*inset*). The off-resonant energies of the two spectral peaks are marked by dashed lines. The excitation energy dependent dispersion and line shape changes of the SXE structures identify effects of momentum conservation in the RIXS process.

channels crucially depend on the ratio of the corresponding matrix elements.)

In view of the anisotropy of GaN, the PDOS and SXE/SXA computations were performed for the N $p_z$ and N $p_{xy}$ orbitals separately. The SXE/SXA spectra relevant for our experimental geometry were obtained by summation of the N $p_z$ and N $p_{xy}$ spectra weighted by relative squared amplitudes of the $\mathbf{E}//\mathbf{c}$ and $\mathbf{E}\perp\mathbf{c}$ components of the $\mathbf{E}$ vector (see below).

The calculated SXE/SXA spectra are shown in Fig.1 in comparison with the experiment. The smooth curves include empirical simulation of the valence hole and conduction electron lifetimes. This was achieved by convolution with a Lorentzian, whose full-width was taken proportional to the energy distance from the Fermi level as $0.1*|E-E_F|$, where the proportionality coefficient was estimated from the experimental spectral broadening. The experimental resolution was also included into this simulation by additional

Gaussian convolution. These spectra were used to set the VB maximum energy position in the experimental spectra by aligning the SXE spectral leading edges.

The SXA computations were further extended to include excitonic effects within the 'final-state approximation'[22] by generation of a self-consistent-field (scf) potential in a supercell with the core hole on a probe atom. These computations employed a 2x2x1 supercell (16 atoms/cell) with one core hole ($1s$) on one of the N atoms. The missing core electron was added to the valence electrons, and during the scf-cycle all states (both of the particular atom with the core hole and of all neighboring atoms) were allowed to respond to this core hole (a larger "effective nuclear-charge") and screen the effect. We did not attempt to "optimize" the theoretical spectra by choosing a partial core hole (e.g. half a $1s$ electron), although it was shown previously that such an approach may lead to even better agreement with experiment.[22] The resulting SXA spectrum is also shown in Fig.1. Further computational results are presented below on the track of discussions.

The theoretical SXE/SXA spectra in Fig.1 demonstrate, on the whole, convincing agreement with the experiment. Even such a subtle detail as the high-energy shoulder of the Ga $3d$ derived peak is well reproduced. Some energy shifts between the spectral structures from theory and experiment can be attributed to the excited-state self-energy corrections, and relative amplitude disagreements to the excitonic effects (see below).

Our computational results are a major improvement compared to the previous LMTO calculations from Ref. 15, which were restricted by the use of the muffin-tin (MT) potential. This illustrates the importance of full-potential effects in covalent materials such as GaN.

## IV. RESULTS AND DISCUSSIONS

### A. SXA and off-resonant SXE

*Anisotropy effects*

In GaN the VB and CB are dominated by the N $2p_z$ and $2p_{xy}$ orbitals oriented, respectively, parallel and perpendicular to the (0001) surface normal **c**. The selection rules in SXA are such that the excitation cross-section into a particular $p$-orbital is maximized if the incident **E** vector is oriented parallel to the orbital axis, and vanishes if perpendicular.[25] Therefore, the **E**//**c** component of the **E** vector excites the $p_z$ orbitals, and the **E**⊥**c** component the $p_{xy}$ ones. Our experimental SXA spectrum is taken near the grazing incidence geometry with a grazing incidence angle of 20°. In this case the ratio between $(\mathbf{E}//\mathbf{c})^2$ and $(\mathbf{E}\perp\mathbf{c})^2$, representing the relative weight of the N $p_z$ and N $p_{xy}$ contributions to the SXA spectrum, is about 7.5:1. Therefore, the SXA spectrum reflects mainly the unoccupied N $p_z$ orbitals. The same selection rules apply to SXE. Our experimental SXE spectrum is taken near the normal emission geometry with an angle of 20° to the surface normal. This yields the ratio between the N $p_z$ and N $p_{xy}$ contributions of about 1:7.5, reverse to the SXA case. Therefore, our SXE spectrum reflects mainly the occupied N $p_{xy}$ orbitals.

In highly anisotropic wurtzite GaN, the PDOS associated with the N $p_z$ and N $p_{xy}$ orbitals is significantly different. This is illustrated by computations in Fig.3. The SXE/SXA spectra are seen to be essentially the PDOS multiplied by the dipole matrix element with weak energy dependence. The difference of the N $p_z$ and N $p_{xy}$ PDOS is thus replicated by differences of the corresponding grazing incidence and normal emission spectra giving rise to the angle dependence of the corresponding SXA and SXE spectra. Dramatic differences of the N $p_z$ and N $p_{xy}$ PDOS in the CB results in a particularly strong angle dependence of the SXA spectra.[9,15] The theoretical SXE/SXA spectra in Fig.1, corresponding to our experimental geometry, were obtained as a superposition of the N $p_z$ and N $p_{xy}$ ones weighted according to the $(\mathbf{E}//\mathbf{c})^2$ to $(\mathbf{E}\perp\mathbf{c})^2$ ratio.

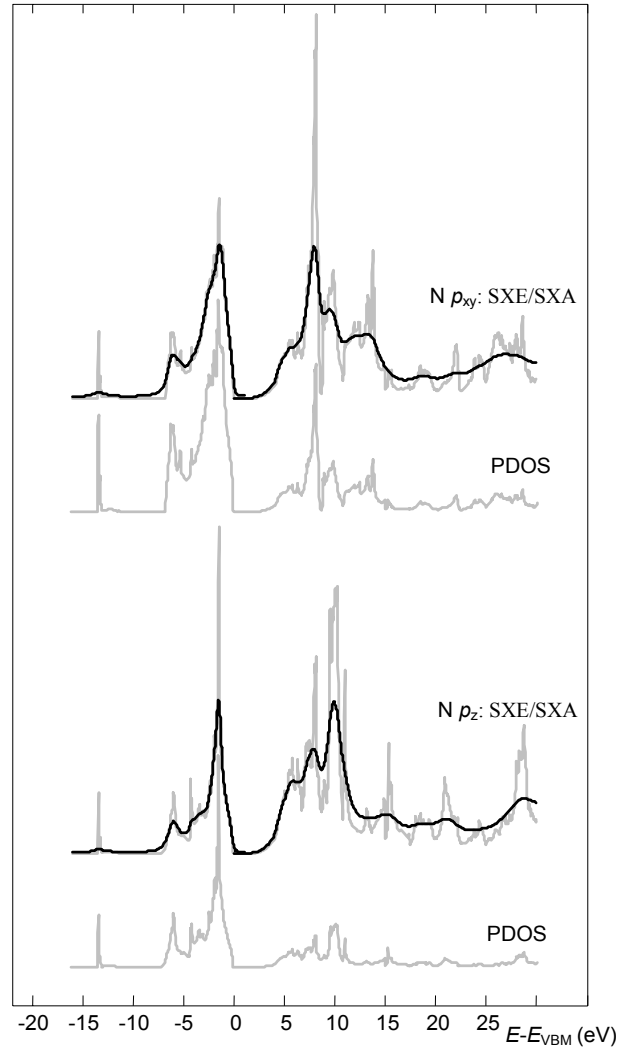

FIG. 3. (*top curves*) N $p_{xy}$ PDOS with the corresponding normal emission SXE and normal incidence SXA spectra, and (*bottom*) N $p_z$ PDOS with the grazing emission and grazing incidence ones. Their differences results in angle dependence of the SXE/SXA spectra. The lifetime and instrumental broadening is simulated as in Fig.1.

*Core hole effects in SXA*

The core hole generated in the SXA process in fact dynamically interacts with the valence electrons and modifies the crystal potential around it (see, e.g., Ref. 26). One extreme static approximation to this process, implied by the above SXA calculations, is the 'initial-state' approximation, where the core hole is completely ignored. Another extreme is the 'final-state' approximation,[23] which considers a crystal potential resulting from static screening of the core hole by valence electrons. It is often argued that the first extreme is more relevant for metals, since they can have a better effective core-hole screening, while the second approach is more relevant for insulators.[22]

The N $p_z$ and N $p_{xy}$ SXA spectra calculated with the screened core hole are shown in Fig.4. In comparison with Fig.3, inclusion of the core hole is seen to change significantly the amplitudes of the spectral structures. A superposition of the N $p_z$ and N $p_{xy}$ spectra corresponding to our experimental geometry is shown in Fig.1. The main change is a dramatic increase of the first SXA peak. While in the no-hole case its amplitude was underestimated compared to the experiment, the core hole effects result in its overestimation. Moreover, the third SXA peak appears reduced compared to the experiment. Incompleteness of both 'initial-state' and 'final-state' approximation suggest partial core hole screening[22] with significant mobility of the valence charge in GaN.

*Excited-state self-energy corrections*

In SXE, if the final-state rule[23] is correct, the spectral structures reflect the quasi-particle energy levels of the valence hole with filled core hole. Their shifts from the DFT energy levels are the self-energy corrections $\Delta\Sigma$, appearing due to difference in the exchange-correlation potential between the excited and ground state. In our results we observe $\Delta\Sigma$ values of almost zero for the dominant SXE peak, -0.4 for the smaller peak, and -4.3 eV for the Ga3$d$ derived peak. As expected, $\Delta\Sigma$ increases when going away from the Fermi level, and is particularly large for the Ga 3$d$ derived peak due to the localized semicore character of this state.[17]
In SXA, the quasi-particle energy levels of conduction electrons are distorted by interaction with the partially screened core hole. If we correct from the interaction energy estimated, very roughly, as half the difference between the SXA calculations with and without the core hole, the $\Delta\Sigma$ shifts for the three leading SXA peaks appear as 0.4, 1.3 and 1.2 eV to higher energies. The sign of the shifts is consistent with the band gap problem of the DFT, which neglects discontinuity of the exchange-correlation for excitations across the band gap. However, their magnitude appears smaller than the $\Delta\Sigma$ shift in the CB minimum (CBM), which is 1.5 eV as determined by the difference between the experimental optical band gap of 3.4 eV and our calculated one of 1.9 eV. In view of the expected $\Delta\Sigma$ increase with energy, such an anomalous behavior of $\Delta\Sigma$ can indicate the self-energy effects depending on the wave function character.[27]

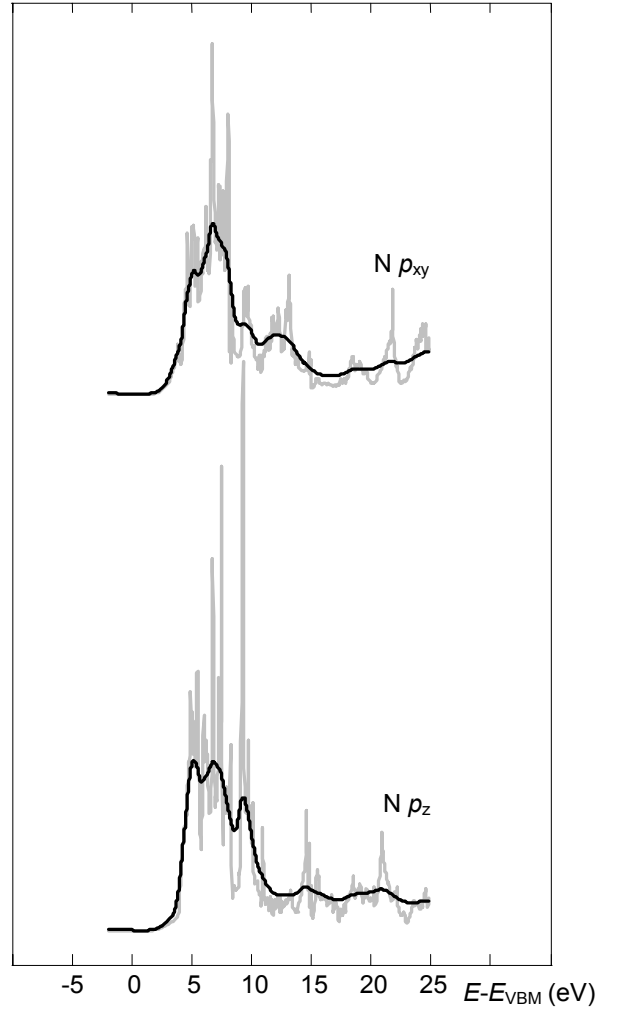

FIG. 4. N $p_z$ and $p_{xy}$ derived SXA spectra similar to Fig.3, but calculated in the 'final-state approximation' with the screened core hole.

**B. RIXS**

*Energy dependence of the coherent fraction*

Changes in the experimental resonant SXE series within the main N 2$p$ line region are emphasized in difference spectra shown in Fig.5 (*top*). These spectra are obtained from the curves in Fig.2 by subtracting the off-resonant reference spectrum. The latter was scaled, for each spectrum, to set the difference spectrum positive and having its minimal value equal to zero. To reduce the influence of noise on this procedure, the scaling coefficient was determined from the spectra denoised by Gaussian smoothing.

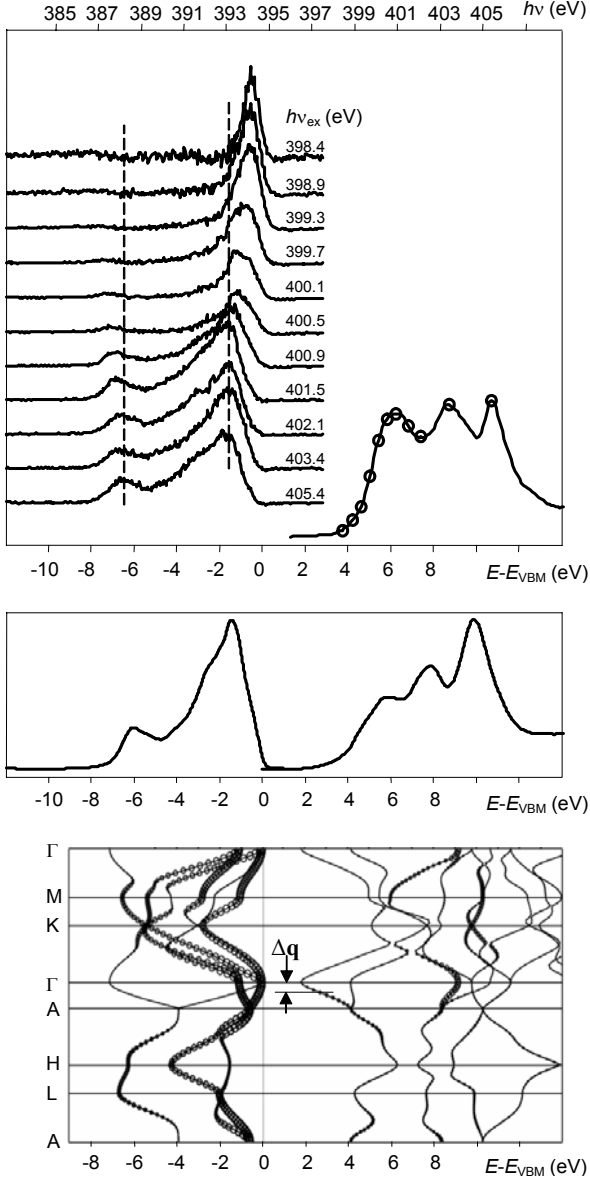

FIG. 5. (*top*) Differential resonant SXE spectra, representing the coherent spectral fraction, normalized to the same intensity maximum. The indicated excitation energies $h\nu_{ex}$ are also marked at the SXA spectrum. The energies of the two off-resonant SXE spectral peaks are marked by dashed lines; (*middle*) calculated N $p_{xy}$ derived SXE and N $p_z$ derived SXA spectra; (*bottom*) calculated $E(\mathbf{k})$. The weights of the N $2p_{xy}$ VB states and N $2p_z$ CB states are indicated by the radii of the circles. Dispersion of the resonant SXE spectral structures is related primarily to that of the N $2p_{xy}$ valence states and N $2p_z$ conduction states.

The difference spectra in Fig.5 represent in some way the coherent fraction in the resonant SXE spectra: If the off-resonant spectrum is – by virtue of averaging over **k**-space due to electron-phonon (*e-ph*) and electron-electron (*e-e*) interactions, and large number of different available CB **k**-states in the intermediate state – the incoherent fraction, the positive intensity remaining after its subtraction is the coherent fraction.[4]

The relative weight of the coherent fraction in the total SXE spectra within the VB as a function of $h\nu_{ex}$ is shown in Fig.6. It was evaluated as the integral over the VB energy interval from 387.4 to 395.4 eV of the difference spectra from Fig.5, divided by the integral of the corresponding total spectra from Fig.2. The $h\nu_{ex}$ dependence of the latter is in fact equivalent to the SXA spectrum, also shown in Fig.6, because the N 1$s$ core hole can be radiatively filled only from the N 2$p$ VB states (the contribution of the Ga 3$d$ derived states is negligible).

The coherent fraction in Fig.6 vanishes at higher energies because the strength of the *e-ph* and *e-e* interactions, driving the scattering process into the incoherent channel, increases with energy due to increase of the phase space available for them. However, below $h\nu_{ex}$ of 405eV our experimental data demonstrate, surprisingly, that the coherent fraction weight behaves non-monotonously and even *increases* with energy. This fact is beyond the above simple phase space arguments. Evidently, the low-energy conduction states in GaN change their character in this energy region in such a way that their susceptibility to the incoherent *e-ph* and *e-e* scattering somehow varies. Tentatively, we attribute this effect to the anisotropy of the material. As the coherent fraction is larger in the region where the CB layer-projected DOS is dominated by N $p_{xy}$ states, it suggests that the incoherent scattering is in some way less pronounced within the atomic layers than between the layers. Interestingly, the coherent fraction remains large up to rather high excitation energies.

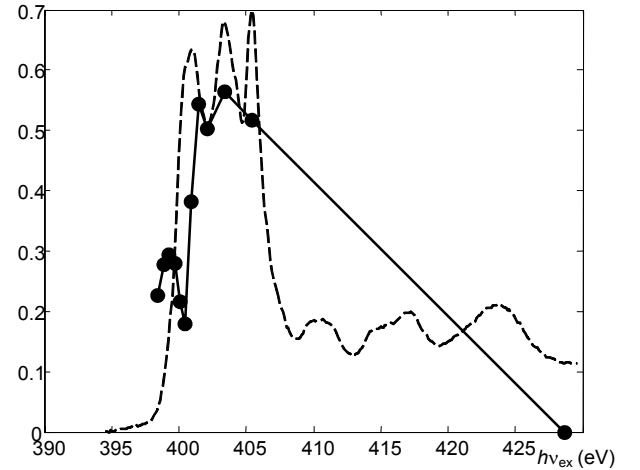

FIG. 6. Relative weight of the coherent SXE fraction as a function of excitation energy (the thin lines connecting the dots are a guide for the eye) compared with the SXA spectrum (dashed line, arbitrarily scaled) reflecting the total SXE intensity. Anomalous behavior of the coherent fraction at low $h\nu_{ex}$ reflects changes in character of the conduction states.

*k–selectivity effects*

Our experimental SXE and SXA spectra should be related, with our incidence and scattering angles, primarily to the bands in $E(\mathbf{k})$ having significant N $p_{xy}$ character in the VB and N $p_z$ in the CB. Fig.5 (*bottom*) shows the calculated $E(\mathbf{k})$ where the corresponding N $2p_{xy}$ weights of the VB states and N $2p_z$ of the CB states are indicated.

As a link between the experimental spectra and calculated $E(\mathbf{k})$, Fig.5 (*middle*) shows the calculated off-resonant N $p_{xy}$ SXE and N $p_z$ SXA spectra from Fig. 3. They are essentially equivalent to the corresponding PDOS. It should be noted that for GaN the peaks in these spectra can not be unambiguously associated with certain critical points along high-symmetry Brillouin zone (BZ) lines. This is because the complicated $E(\mathbf{k})$ of GaN results in a multitude of critical points, including those from non-symmetry BZ directions, whose energy separation is smaller than the lifetime broadening.

A few k-selectivity effects can nevertheless be unambiguously identified in evolution of the resonant SXE spectra with $hv_{ex}$ (as emphasized in the difference spectra) in comparison with the calculated $E(\mathbf{k})$:

(1) most remarkably, tuning $hv_{ex}$ towards the absorption onset at 398.6 eV results in shifting of the dominant SXE peak to higher energies, with an additional peak splitting off on its high energy side. As seen from the comparison of the calculated SXA spectrum to the N $p_z$ band structure, for the absorption onset energy the excited electrons appear in the CB already in some 1 eV above the CBM (such a delayed SXA onset occurs, as can be seen in the $E(\mathbf{k})$ plot, because the lowest CB states, depleted in the N $p$ character and having rather steep dispersions, yield only a slight tail in the PDOS and thus in the SXA spectrum). The corresponding **k**-vectors distribute near ΓA in some 0.3 Å$^{-1}$ from the Γ-point. The photon wavevector transfer $\Delta\mathbf{q}=\mathbf{q}_{out}-\mathbf{q}_{in}$ in our experiment has almost the same value 0.28 Å$^{-1}$, and is also directed close to ΓA. Therefore, as indicated in the $E(\mathbf{k})$ panel, the k-conserving RIXS process couples the excited N $p_z$ conduction electrons to the N $p_{xy}$ valence holes in the VB maximum in the Γ-point, which results in a strong coherent emission from the VB maximum appearing on the high-energy side of the main incoherent peak. In the experimental resonant series this is seen as shifting of the SXE spectral maximum to higher energies, with the coherent emission at the absorption threshold splitting off in a separate peak.

Tuning $hv_{ex}$ below 398.2eV can be expected to shift the conduction electron closer to the Γ-point, whereas the valence hole is shifted away from this point. This should result in shifting of the coherent peak again to lower energies. However, very low absorption in this $hv_{ex}$ region makes such experiment difficult.

The VB minimum is placed in the same Γ-point as the VBM. Therefore, it also shows an increase of the coherent SXE intensity on the absorption onset, most clear in the difference spectra as an intensity enhancement near a binding energy of -8.2 eV. Taking into account the self-energy effects, this figure is consistent with the theoretical energy -7.1 eV.

It should be noted that due to relatively high photon energies at the N *K*-edge the proper interpretation of RIXS requires taking into account the photon momentum transfer. Also the GaN anisotropy effects become crucial, with the orbital orientation selectivity allowing us to disentangle the **k**-dispersions of various states;

(2) with increase of $hv_{ex}$ the **k**-vector, following the $2p_z$ conduction band, moves away from the Γ-point along the ΓAH line. Correspondingly, the dominant SXE peak disperses to lower energies, reflecting the $2p_{xy}$ valence band dispersions along ΓAH. Its dispersion range towards $hv_{ex}$ of 401eV, where the peak appears near the incoherent energy position, is around 0.9 eV. With the calculated CB dispersions, this figure well matches the dispersion of the $2p_{xy}$ heavy-hole valence bands. We do not detect any dispersion renormalization due to coupling to the core hole. Any contribution of the light-hole valence band is not seen in the SXE spectra because of its $p_z$ character in this region of the **k**-space;

(3) with $hv_{ex}$ varied near 401 eV, the SXE spectra display a sizeable dispersing coherent fraction below the smaller incoherent peak at higher binding energies. Based on the N $p_z$ dispersions in the CB, it can be associated with the N $p_{xy}$ dispersions in the M-valley near the VB bottom;

(4) upon further increase of $hv_{ex}$ above 401 eV the coherent peak moves to the other side of the incoherent one. Its energy position can be associated with an integral effect of a few VB critical points in ~2eV below the VBM such as that in the M-point.

*Emission angle dependence*

The above discussion involving the anisotropy effects suggests that the resonant SXE spectra should show certain dependence on the emission angle. To check this, we rotated the sample towards the normal incidence and grazing emission geometry with a grazing emission angle of 20°, opposite to the experimental geometry above. In this case the absorption is mainly due to the unoccupied N $p_{xy}$ orbitals and emission due to the occupied N $p_z$ ones. However, the SXE signal in this geometry gets strongly suppressed due to the self-absorption. We have therefore measured only one representative resonant SXE spectrum with $hv_{ex}$=399.3 eV not far from the absorption threshold.

The experimental grazing emission spectrum is shown in Fig.7 compared with the normal emission one from Fig.2 measured at the same $hv_{ex}$. Their difference is plotted below. The spectra show statistically significant differences, reflecting the difference between the N $p_{xy}$ and N $p_z$ states. In particular, in the grazing emission spectrum the spectral maximum is shifted by ~0.15 eV to deeper energies. This reflects more steep dispersion of the valence N $p_z$ states away from the Γ-point compared to the N $p_{xy}$ states.

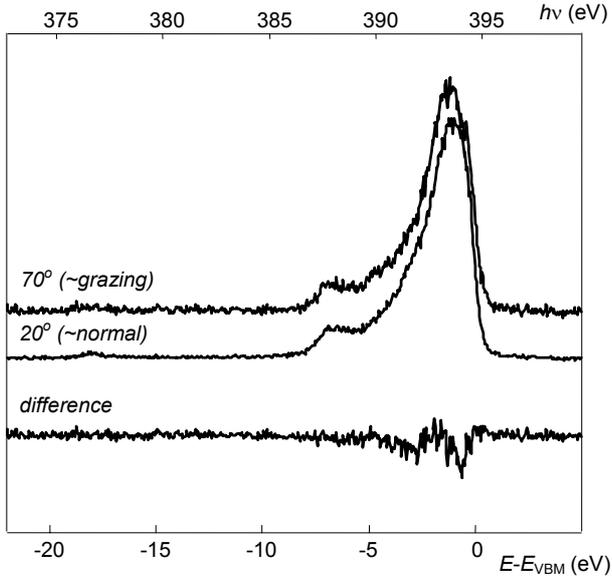

FIG.7. Resonant SXE spectrum measured with $h\nu_{ex}$=399.3 eV near the grazing emission geometry (*top curve*) compared to its normal emission counterpart from Fig.2 normalized to the same peak value (*below*), and their difference (*bottom*). The difference between the spectra is due to the difference between the N $p_{xy}$ and N $p_z$ states.

## V. SUMMARY AND CONCLUSIONS

High-resolution SXE/SXA experimental data on wurtzite GaN near the N *K*-edge are presented. The measurements are supported by full-potential calculations extended to the core hole screening.

The obtained results identify, in particular: (1) partial core hole screening in the SXA process; (2) effects of the GaN anisotropy in the SXE/SXA processes. Our experimental geometry invoked primarily the $p_{xy}$ states in the VB and $p_z$ states in the CB; (3) pronounced dispersions of the resonant SXE structures, identifying the effects of momentum conservation and **k**-selectivity in the RIXS process; (4) non-monotonous behavior of the coherent SXE fraction, reflecting different effects of the involved CB states in the *e-e* and *e-ph* scattering.

In view of a wide range of optoelectronic applications of GaN, especially in the short wavelength region, our findings on the **k**-selectivity and anisotropy effects in RIXS can be utilized to guide future development in GaN based nanostructure devices by controlling direct *vs* indirect band gap character and crystallographic orientation of GaN nanostructures.

anomalous strengths of the non-dipole transitions, whose origin would then be unclear.